\documentclass[prd,aps,showpacs,onecolumn]{revtex4}
\usepackage{amssymb}
\usepackage{amsfonts}
\usepackage{amsmath}
\usepackage{graphicx}
\usepackage{bm}

\begin{document}

\title{New rational extensions of solvable potentials with finite bound
state spectrum}
\author{Yves Grandati }
\affiliation{Institut de Physique, Equipe BioPhyStat, ICPMB, IF CNRS 2843, Universit\'{e}
Paul Verlaine-Metz, 1 Bd Arago, 57078 Metz, Cedex 3, France}

\begin{abstract}
Using the disconjugacy properties of the Schr\"{o}dinger equation, it is
possible to develop a new type of generalized SUSY\ QM partnership which
allows to generate new solvable rational extensions for translationally
shape invariant potentials having a finite bound state spectrum.

For this we prolong the dispersion relation relating the energy to the
quantum number out of the physical domain until a disconjugacy sector. The
prolonged excited states Riccati-Schr\"{o}dinger (RS) functions are used to
build Darboux-B\"{a}cklund transforms which give regular isospectral
extensions of the initial potential. We give the spectra of these extensions
in terms of new orthogonal polynomials and study their shape invariance
properties.
\end{abstract}

\maketitle

\section{\protect\bigskip Introduction}

The last three years have seen a substantial development of research works
concerning the study of rational extensions of solvable quantum potentials,
in particular because of their intimate relation with the recently
discovered exceptional orthogonal polynomials (EOP) \cite%
{gomez,gomez2,gomez3,gomez4,gomez5,gomez6,gomez7,quesne1,quesne,quesne2,quesne3,quesne4,quesne5,odake,sasaki,ho,odake2,odake3,sasaki2,dutta,grandati2,grandati3,grandati4,grandati5}%
.

In \cite{grandati2,grandati3,grandati4,grandati5} we have proposed a new
scheme to generate the rational extensions of every primary translationally
shape-invariant potentials (PTSIP) \cite{cooper,Dutt,Gendenshtein,grandati}.
They are obtained via Darboux-B\"{a}cklund Transformations (DBT) based on
negative eigenfunctions built from excited states of the initial
hamiltonian. The regularity of corresponding Riccati-Schr\"{o}dinger (RS)
functions \cite{grandati} is then directly verified by combining
disconjugacy theorems and asymptotic analysis of the eigenfunctions. This
approach which is systematic and generalizes the usual SUSY QM partnership
can also be enlarged in a multistep version. In \cite%
{grandati2,grandati3,grandati4,grandati5} these unphysical eigenfunctions
are deduced from the excited states via discrete symmetries acting on the
set of parameters of the initial potential. For the isotonic or
trigonometric Darboux-P\"{o}schl-Teller potentials, when the obtained
extensions are strictly isospectral to the original potential, their
eigenstates are (up to a gauge factor) the EOP discovered by Gomez-Ullate,
Kamran and Milson \cite{gomez}.

For potentials presenting an infinite number of bound states associated to
all integer values of the quantum number, the recourse to discrete
symmetries is necessary to reach the disconjugacy sectors. However for PTSIP
with a finite bound state spectrum $\left\{ E_{n},\ n\leq n_{\max }\right\} $
there exists another way. In this case, some disconjugacy sectors can be
attained by prolonging the eigenstates $\psi _{n}$ for values of the quantum
number $n$ going beyond $n_{\max }$. If the "dispersion relation", ie $E_{n}$
as a function of $n$, goes then to negative values, the corresponding
prolonged eigenstates $\psi _{n}$, although diverging at least at one
extremity of the definition interval, may possibly be used to build regular
extensions via DBT. Here again, disconjugacy theorems and asymptotic
analysis allow to control the regularity of the corresponding RS\ function $%
w_{n}$. In the present paper we consider the four PTSIP for which the
associated dispersion relation has the mentioned behaviour. We can share
them into two groups: the Morse and hyperbolic Darboux-P\"{o}schl-Teller
potentials, which have a parabolic (quadratic) dispersion relation onb the
one hand and the Eckart and hyperbolic Rosen-Morse (HRM) potential for which
the dispersion relation has the form of a second degree Laurent polynomial
on the other hand. For the first group the prongation leads to only one
disconjugacy sector and the extensions obtained are strictly isosectral to
the initial potential. For the Morse potential we recover exactly the new
extended potentials obtained very recently by Quesne \cite{quesne6}. Quesne
shows in particular that if these extensions don't inherit of the
translational shape invariance of the original potential (which one of the
feature of the extensions associated to the EOP), they satisfy a kind of
"enlarged shape invariance property". We prove that it is also the case for
the similar extensions of the HDPT potential. \ For the Eckart potential,
again we have only one disconjugacy sector with strictly isospectral
extensions. This is no more the case for the HRM potential. In this last
case, we have three distinct disconjugacy sectors, the two first
corresponding to strictly isospectral extensions while the DBT built on
eigenfunctions of the third sector are reverse SUSY QM partnership and then
give only quasi isospectral extensions. For all the obtained strictly
isospectral extensions, we give explicit expressions for the eigenstates in
terms of new orthogonal polynomials.

\section{Disconjugacy and regular extensions of one dimensional potentials}

If $\psi _{\lambda }(x;a)$ is an eigenstate of $\widehat{H}%
(a)=-d^{2}/dx^{2}+V(x;a),\ a\in \mathbb{R}^{m},\ x\in I\subset \mathbb{R},$
associated to the eigenvalue $E_{\lambda }(a)$ ($E_{0}(a)=0$)

\begin{equation}
\psi _{\lambda }^{\prime \prime }(x;a)+\left( E_{\lambda }(a)-V(x;a)\right)
\psi _{\lambda }(x;a)=0,  \label{EdS}
\end{equation}
then the Riccati-Schr\"{o}dinger (RS) function $w_{\lambda }(x;a)=-\psi
_{\lambda }^{\prime }(x;a)/\psi _{\lambda }(x;a)$ satisfies the
corresponding Riccati-Schr\"{o}dinger (RS) equation \cite{grandati}

\begin{equation}
-w_{\lambda }^{\prime }(x;a)+w_{\lambda }^{2}(x;a)=V(x;a)-E_{\lambda }(a).
\label{edr4}
\end{equation}

The set of Riccati-Schr\"{o}dinger equations is preserved by the Darboux-B%
\"{a}cklund Transformations (DBT), which are built from any solution $w_{\nu
}(x;a)$ of the initial RS equation Eq(\ref{edr4}) as \cite%
{carinena2,Ramos,grandati}

\begin{equation}
w_{\lambda }(x;a)\overset{A\left( w_{\nu }\right) }{\rightarrow }w_{\lambda
}^{\left( \nu \right) }(x;a)=-w_{\nu }(x;a)+\frac{E_{\lambda }(a)-E_{\nu }(a)%
}{w_{\nu }(x;a)-w_{\lambda }(x;a)},  \label{transfoback2}
\end{equation}%
where $E_{\lambda }(a)>E_{\nu }(a)$. $w_{\lambda }^{\left( \nu \right) }$ is
then a solution of the RS equation:

\begin{equation}
-w_{\lambda }^{\left( \nu \right) \prime }(x;a)+\left( w_{\lambda }^{(\nu
)}(x;a)\right) ^{2}=V^{\left( \nu \right) }(x;a)-E_{\lambda }(a),
\label{eqtransform}
\end{equation}%
with the same energy $E_{\lambda }(a)$ as in Eq(\ref{edr4}) but with a
modified potential

\begin{equation}
V^{\left( \nu \right) }(x;a)=V(x;a)+2w_{\nu }^{\prime }(x;a),
\label{pottrans}
\end{equation}%
which is called an extension of $V$.

The corresponding eigenstate of $\widehat{H}^{\left( \nu \right)
}(a)=-d^{2}/dx^{2}+V^{\left( \nu \right) }(x;a)$ can be written

\begin{equation}
\psi _{\lambda }^{\left( \nu \right) }(x;a)=\exp \left( -\int dxw_{\lambda
}^{(\nu )}(x;a)\right) \sim \frac{1}{\sqrt{E_{\lambda }\left( a\right)
-E_{\nu }(a)}}\widehat{A}\left( w_{\nu }\right) \psi _{\lambda }(x;a),
\label{foDBT}
\end{equation}%
where $\widehat{A}\left( a\right) $ is a first order operator given by

\begin{equation}
\widehat{A}\left( w_{\nu }\right) =d/dx+w_{\nu }(x;a).  \label{opA}
\end{equation}

From $V$, the DBT generates a new potential $V^{\left( \nu \right) }$
(quasi)isospectral to the original one and its eigenfunctions are directly
obtained from those of $V$ via Eq(\ref{foDBT}). Nevertheless, in general, $%
w_{\nu }(x;a)$ and the transformed potential $V^{\left( \nu \right) }(x;a)$
are singular at the nodes of $\psi _{\nu }(x;a)$. For instance, if $\psi
_{n}(x;a)$ ($\nu =n$) is a bound state of $\widehat{H}(a)$, $V^{\left(
n\right) }$ is regular only when $n=0$, that is when $\psi _{n=0}$ is the
ground state of $\widehat{H}$, and we recover the usual SUSY partnership in
quantum mechanics.

We can however envisage to use any other regular solution of Eq(\ref{edr4})
as long as it has no zero on the considered real interval $I$, even if it
does not correspond to a physical state. In particular, it is possible to
use some eigenfunctions $\psi _{\nu }$ associated to negative eigenvalues $%
E_{\nu }(a)<0$ \cite{sukumar}. This is due to the disconjugacy of the Schr%
\"{o}dinger equation for these eigenvalues \cite{grandati5}. More precisely,
a second order differential equation like Eq(\ref{EdS}) is said to be
disconjugated on $I$ if every solution of this equation has at most one zero
on $I$ \cite{hartman,coppel,derr}. For a closed or open interval $I$, the
disconjugacy of Eq(\ref{EdS}) is equivalent to the existence of solutions of
this equation which are everywhere non zero on $I$ \cite{hartman,coppel,derr}%
.

We have also the following result

\emph{Theorem} \cite{hartman,coppel}\ If there exists a continuously
differentiable solution on $I$\ of the Riccati inequation

\begin{equation}
-w^{\prime }(x)+w^{2}(x)+G(x)\leq 0  \label{Ricineq}
\end{equation}%
then the equation

\begin{equation}
\psi ^{\prime \prime }(x)+G(x)\psi (x)=0,  \label{EdSb}
\end{equation}%
is disconjugated on $I$.

\ \ \ \ In our case, if $E_{\lambda }(a)\leq 0$, we have

\begin{equation}
-w_{0}^{\prime }(x;a)+w_{0}^{2}(x;a)=V(x;a)\leq V(x;a)-E_{\lambda }(a),
\end{equation}%
$w_{0}(x;a)$ being continuously differentiable on $I$. The above theorem
ensures the existence of nodeless solutions $\phi _{\lambda }(x;a)$ of Eq(%
\ref{EdS}), that is, of corresponding regular RS functions $v_{\lambda
}(x;a)=-\phi _{\lambda }^{\prime }(x;a)/\phi _{\lambda }(x;a)$. To prove
that a given solution $\phi _{\lambda }(x;a)$ belongs to this category, it
is sufficient to determine the signs of the limit values $\phi _{\lambda
}(0^{+};a)$ and $\phi _{\lambda }(+\infty ;a)$. If they are identical then $%
\phi _{\lambda }$ is nodeless and if they are opposite $\phi _{\lambda }$
presents then a unique zero on $I$. In the first case $V(x;a)+2v_{\lambda
}^{\prime }(x;a)$ constitutes a perfectly regular (quasi)isospectral
extension of $V(x;a)$.\ 

Of particular interest is to obtain such solutions $\phi _{\lambda }(x;a)$
which are polynomials (up to a gauge factor) in order to build rational
extensions of the initial potential $V$. In \cite%
{grandati3,grandati4,grandati5}, e have shown that it is possible to
generate such solutions by using specific discrete symmetries $\Gamma _{i}$
which are covariance transformations for the considered family of potentials

\begin{equation}
\left\{ 
\begin{array}{c}
a\overset{\Gamma _{i}}{\rightarrow }a_{i} \\ 
V(x;a)\overset{\Gamma _{i}}{\rightarrow }V(x;a_{i})=V(x;a)+U\left( a\right) .%
\end{array}%
\right.  \label{sym}
\end{equation}

$\Gamma _{i}$ acts on the parameters of the potential and transforms the RS
function of a\ physical excited eigenstate $w_{n}$ into a unphysical RS
function $v_{n,i}(x;a)=\Gamma _{i}\left( w_{n}(x;a)\right) =w_{n}(x;a_{i})$
associated to the negative eigenvalue $\mathcal{E}_{n,i}(a)=\Gamma
_{i}\left( E_{n}(a)\right) =U\left( a\right) -E_{n}(a_{i})<0$.

\begin{equation}
-v_{n,i}^{\prime }(x;a)+v_{n,i}^{2}(x;a)=V(x;a)-\mathcal{E}_{n,i}(a).
\label{EqRSpourv}
\end{equation}

To $v_{n,i}$ corresponds an unphysical eigenfunction of $\widehat{H}(a)$

\begin{equation}
\phi _{n,i}(x;a)=\exp \left( -\int dxv_{n,i}(x;a)\right)  \label{phi}
\end{equation}%
associated to the eigenvalue $\mathcal{E}_{n,i}(a)$.

Since $\mathcal{E}_{n,i}(a)<0$, if $\phi _{n,i}(x;a)$ has the same sign at
both extremities of $I$, then $v_{n,i}(x;a)$ can be used to build a regular
extended potential (see Eq(\ref{pottrans}) and Eq(\ref{foDBT}))

\begin{equation}
V^{\left( n,i\right) }(x;a)=V(x;a)+2v_{n,i}^{\prime }(x;a)
\end{equation}%
(quasi)isospectral to $V(x;a)$. The eigenstates of $V^{\left( n,i\right) }$
are given by (see Eq(\ref{transfoback2}))

\begin{equation}
\left\{ 
\begin{array}{c}
w_{k}^{\left( n,i\right) }(x;a)=-v_{n,i}(x;a)+\frac{E_{k}(a)-\mathcal{E}%
_{n,i}(a)}{v_{n,i}(x;a)-w_{k}(x;a)} \\ 
\psi _{k}^{\left( n,i\right) }(x;a)=\exp \left( -\int
dxw_{k}^{(n,i)}(x;a)\right) \sim \widehat{A}\left( v_{n,i}\right) \psi
_{k}(x;a)%
\end{array}%
\right. ,  \label{foext}
\end{equation}%
for the respective\ energies $E_{k}(a)$.\ 

The nature of the isospectrality depends if $1/\phi _{n,i}(x;a)$\ satisfies
or not the appropriate boundary conditions. If it is the case, then $1/\phi
_{n,i}(x;a)$ is a physical eigenstate of $\widehat{H}^{\left( n,i\right)
}(a)=-d^{2}/dx^{2}+V^{\left( n,i\right) }(x;a)$ for the eigenvalue $\mathcal{%
E}_{n,i}(a)$ and we only have quasi-isospectrality between $V(x;a)$\ and\ $%
V^{\left( n,i\right) }(x;a)$. If it is not the case, the isospectrality
between $V^{\left( n,i\right) }(x;a)$ and $V(x;a)$ is strict.

This procedure can be viewed as a "generalized SUSY\ QM\ partnership" where
the DBT can be based on excited states RS\ functions properly regularized by
the symmetry $\Gamma _{j}$. In \cite{grandati3,grandati4,grandati5}, it has
been applied to exceptional PTSIP of the first and second categories \cite%
{grandati}. In the particular case of the isotonic oscillator, the spectrum
of the two first series of extensions are given (up to a gauge factor) by
the exceptional Laguerre polynomials initially discovered by Gomez-Ullate et
al \cite{gomez3}.

For some potentials which have finite bound states spectrum, it is possible
to use an even more direct way to generate negative energies eigenstates. In
this case, the eigenfunctions $\psi _{n}(x;a)$ satisfy the required
Dirichlet boundary conditions to be acceptable physical eigenstates only for
a finite number of values $\left\{ 0,...,n_{\max }\right\} $ of the quantum
number $n$. Beyond this maximal value $n_{\max }$, $\psi _{n}$ have a
divergent behaviour at (at least) one extremity of the definition interval $%
I $. $\psi _{n}$ has then to be rejected as eigenstate but it can still be
used to build the corresponding DBT $A\left( w_{n}\right) $, the energy $%
E_{n}$ (viewed as a function of the quantum number $n$) being extended to
values of $n$ greater than $n_{\max }$. If in this extended domain $E_{n}$
reaches negative values we recover then a disconjugacy sector of the Schr%
\"{o}dinger equation and $\psi _{n}$ can be exempt of nodes. The DBT $%
A\left( w_{n}\right) $ gives then a regular extended potential (see Eq(\ref%
{pottrans}) and Eq(\ref{foDBT}))

\begin{equation}
V^{\left( n\right) }(x;a)=V(x;a)+2w_{n}^{\prime }(x;a),  \label{extpot}
\end{equation}%
(quasi)isospectral to $V(x;a)$ and its eigenstates are given by (see Eq(\ref%
{transfoback2}))

\begin{equation}
\left\{ 
\begin{array}{c}
w_{k}^{\left( n\right) }(x;a)=-w_{n}(x;a)+\frac{E_{k}(a)-E_{n}(a)}{%
w_{n}(x;a)-w_{k}(x;a)} \\ 
\psi _{k}^{\left( n\right) }(x;a)=\exp \left( -\int
dxw_{k}^{(n)}(x;a)\right) \sim \widehat{A}\left( w_{n}\right) \psi _{k}(x;a).%
\end{array}%
\right.  \label{specextpot}
\end{equation}

\section{Primary translationally shape invariant potentials with finite boud
states spectrum}

In \cite{grandati}, we have shown that all the PTSIP can be classified into
two categories in which the potentials can be brought into a harmonic or
isotonic form respectively, using a change of variable which satisfies a
constant coefficient Riccati equation. Among them we have potentials
admitting an infinite bound state spectrum \cite{cooper,Dutt,grandati}. This
is naturally the case of the confining potentials as the harmonic and
isotonic oscillators, the trigonometric DPT (or Scarf I) and the
trigonometric Rosen-Morse potentials. For these ones the dispersion relation
is a strictly increasing function on $n\in \left[ 0,+\infty \right[ $ which
is linear for the two first, parabolic for the third one and which is a
second degree Laurent polynomial for the fourth one. This is also the case
of the effective radial Kepler-Coulomb (ERKC) potential which admits an
infinite bound state spectrum with negative energies and a continuous
spectrum of scattering states with positive energies. The dispersion
relation for the bound states (a second degree Laurent polynomial without a
regular term) is also strictly increasing for $n\in \left[ 0,+\infty \right[ 
$.

The other primary TSIP have a finite bound state spectrum. This is the case
in the first category for the Morse, Eckart and hyperbolic Rosen-Morse
potentials and in the second category for the hyperbolic Darboux-P\"{o}%
schl-Teller (or\ Scarf II) potential \cite{cooper,Dutt,grandati}.

\subsection{Case of dispersion relation which is a second degree polynomial}

\subsubsection{Morse\protect\bigskip\ potential}

For the Morse\bigskip\ potential \cite{cooper,Dutt,grandati} ($x\in \mathbb{R%
}$)%
\begin{equation}
V(y;a,b)=b^{2}y^{2}-2\left( a+\frac{1}{2}\right) by+a^{2},\ a,b>0,
\label{potmorse}
\end{equation}%
where$\ y=\exp \left( -x\right) $, we have a "dispersion relation" (energy $%
E_{n}$ as a function of the quantum number $n\geq 0$) which is of parabolic
type. Namely, we have ($a_{n}=a-n$)

\begin{equation}
E_{n}\left( a\right) =a^{2}-a_{n}^{2}=n\left( 2a-n\right) .
\label{spectremorse}
\end{equation}

The bound states are obtained on the increasing part of $E_{n}$, that is for 
$n<a$ and we have exactly $\left[ a\right] $ bound states ( $\left[ a\right] 
$ being the integer part of $a$) given by

\begin{equation}
\psi _{n}\left( x;a,b\right) =\psi _{0}\left( x;a_{n},b\right) \mathit{L}%
_{n}^{2a_{n}}(2by),  \label{foMorse}
\end{equation}%
where

\begin{equation}
\psi _{0}\left( x;a,b\right) =y^{a}e^{-by}.  \label{fondmorse}
\end{equation}

The corresponding RS functions are given by

\begin{equation}
w_{n}\left( x;a,b\right) =w_{0}\left( x;a_{n},b\right) -2by\frac{\mathit{L}%
_{n-1}^{2a_{n}+1}(2by)}{\mathit{L}_{n}^{2a_{n}}(2by)},  \label{RSfonctmorse}
\end{equation}%
where

\begin{equation}
w_{0}\left( x;a,b\right) =a-by.  \label{RSfondmorse}
\end{equation}

Beyond $n=a$, the $\psi _{n}$ no longer correspond to physical eigenstates
and the $E_{n}$ function decreases. When $n$ exceeds the value $2a$ , $E_{n}$
becomes negative and the corresponding Schr\"{o}dinger equation enters in a
disconjugacy regime.

Using

\begin{equation}
\left\{ 
\begin{array}{c}
\mathit{L}_{n}^{\alpha }\left( z\right) \underset{x\rightarrow 0^{+}}{%
\rightarrow }\frac{\left( \alpha +1\right) _{n}}{n!} \\ 
\mathit{L}_{n}^{\alpha }\left( z\right) \underset{x\rightarrow +\infty }{%
\sim }\frac{(-1)^{n}}{n!}z^{n}%
\end{array}%
\right. ,  \label{limitlaguerre}
\end{equation}%
where $\left( X\right) _{n}=\left( X\right) ...\left( X+n-1\right) $ is the
usual Pochhammer symbol \cite{magnus}, we find that the asymptotic behaviour
of $\psi _{n}\left( x;a,b\right) $ at $+\infty $ and $-\infty $ is given by

\begin{equation}
\psi _{n}\left( x;a,b\right) \underset{x\rightarrow +\infty }{\sim }y^{n-a}%
\frac{\left( 2a-2n+1\right) ...\left( 2a-n\right) }{n!}\underset{%
x\rightarrow +\infty }{\rightarrow }\pm \infty ,
\end{equation}%
and

\begin{equation}
\psi _{n}\left( x;a,b\right) \underset{x\rightarrow -\infty }{\sim }\frac{%
(-1)^{n}}{n!}y^{a}e^{-by}\underset{x\rightarrow +\infty }{\rightarrow }%
0^{\pm },
\end{equation}%
where $\pm \ =\left( -1\right) ^{n}$.

We see that $\psi _{n}$ has the same sign at both ends of the definition
interval, the disconjugacy of the Schr\"{o}dinger equation that it satisfies
implies that $\psi _{n}$ has no zero on $\mathbb{R}$. We can then use the
corresponding RS functions to build (quasi)isospectral extensions of the
Morse potential respectively. They are given by

\begin{equation}
V^{\left( n\right) }(x;a,b)=V(y;a,b)-2y\frac{dw_{n}\left( y;a_{n},b\right) }{%
dy},
\end{equation}%
where $w_{n}\left( y;a_{n},b\right) $ is given in Eq(\ref{RSfonctmorse}),
with $\ y=\exp (-\alpha x)$.

Since $1/\psi _{n}(x;a,b)$ is divergent at $-\infty $, it is not a physical
eigenstate of $V^{\left( n\right) }$ which is then strictly isospectral to $%
V $ and admits for bound state eigenfunctions ($k\in \left\{ 0,...,\left[ a-1%
\right] \right\} $, $z=2by$)

\begin{equation}
\psi _{k}^{\left( n\right) }(x;a,b)=\left( w_{n}(x;a,b)-w_{k}(x;a,b)\right)
\psi _{k}(x;a,b)=\psi _{0}\left( x;a_{k},b\right) \frac{\mathit{L}%
_{n,k}^{a}(z)}{\mathit{L}_{n}^{2a_{n}}(z)},
\end{equation}%
where, making use of the following recurrency properties of the Laguerre
polynomials \cite{szego,magnus}

\begin{equation}
\left\{ 
\begin{array}{c}
\mathit{L}_{n}^{\left( \alpha \right) }\left( z\right) +\mathit{L}%
_{n-1}^{\left( \alpha +1\right) }\left( z\right) =\mathit{L}_{n}^{\left(
\alpha +1\right) }\left( z\right) \\ 
z\mathit{L}_{n-1}^{\alpha +1}\left( z\right) =\left( n+\alpha \right) 
\mathit{L}_{n-1}^{\alpha }\left( z\right) -n\mathit{L}_{n}^{\alpha }\left(
z\right) ,%
\end{array}%
\right.
\end{equation}%
we can write

\begin{equation}
\mathit{L}_{n,k}^{a}(z)=\left( 2a-k\right) \mathit{L}_{n}^{2a_{n}}(z)\mathit{%
L}_{k-1}^{2a_{k}}(z)-\left( 2a-n\right) \mathit{L}_{k}^{2a_{k}}(z)\mathit{L}%
_{n-1}^{2a_{n}}(z),  \label{OPlaguerreetendu}
\end{equation}%
which is a polynomial of degree $n+k-1$.

\subsubsection{HDPT potential}

The dispersion relation of the hyperbolic Darboux-P\"{o}schl-Teller (HDPT)
potential \cite{cooper,Dutt,grandati} ($x>0$)%
\begin{equation}
V(x;\alpha ,\beta )=\frac{(\alpha +1/2)(\alpha -1/2)}{\sinh ^{2}x}-\frac{%
(\beta +1/2)(\beta -1/2)}{\cosh ^{2}x}+V_{0}(\alpha ,\beta ),\ \ \beta
>\alpha +1>1/2,  \label{potHDPT}
\end{equation}%
with $V_{0}(\alpha ,\beta )=(\beta -\alpha -1)^{2}$, is of the same
parabolic type than in the preceding case

\begin{equation}
E_{n}\left( \alpha ,\beta \right) =4n\left( 2a-n\right) ,\text{ }
\label{specHDPT}
\end{equation}%
with $a=\left( \beta -\alpha -1\right) /2$. Here again the bound states are
obtained on the increasing part of $E_{n}$, that is, for $n<a$ and they are
given by%
\begin{equation}
\psi _{n}\left( x;\alpha ,\beta \right) =\psi _{0}\left( x;\alpha ,\beta
\right) \mathit{P}_{n}^{\left( \alpha ,-\beta \right) }\left( z\right) ,
\label{foHDPT}
\end{equation}%
where $z=\cosh 2x$ and

\begin{equation}
\psi _{0}\left( x;\alpha ,\beta \right) =\left( \sinh x\right) ^{\alpha
+1/2}\left( \cosh x\right) ^{-\beta +1/2}.  \label{fondDPT}
\end{equation}

Using \cite{szego,magnus}

\begin{equation}
\left( \mathit{P}_{n}^{\left( \alpha ,\beta \right) }\left( x\right) \right)
^{\prime }=\frac{n+\alpha +\beta +1}{2}\mathit{P}_{n-1}^{\left( \alpha
+1,\beta +1\right) }\left( x\right) ,  \label{derivjacobi}
\end{equation}%
we obtain for the corresponding RS functions

\begin{equation}
w_{n}\left( x;\alpha ,\beta \right) =w_{0}\left( x;\alpha ,\beta \right)
-\sinh 2x\left( n+\alpha -\beta +1\right) \frac{\mathit{P}_{n-1}^{\left(
\alpha +1,1-\beta \right) }\left( z\right) }{\mathit{P}_{n}^{\left( \alpha
,-\beta \right) }\left( z\right) },  \label{RSHDPT}
\end{equation}%
where 
\begin{equation}
w_{0}(x;\alpha ,\beta )=-\left( \alpha +1/2\right) \coth x+\left( \beta
-1/2\right) \tanh x.  \label{RSfondDPT}
\end{equation}

As before, when $n$ exceeds the value $2a$, the corresponding Schr\"{o}%
dinger equation enters in a disconjugacy regime. Using

\begin{equation}
\left\{ 
\begin{array}{c}
\mathit{P}_{n}^{\left( \alpha ,\beta \right) }\left( 1\right) =\binom{%
n+\alpha }{n}=\frac{\left( \alpha +1\right) _{n}}{n!} \\ 
\mathit{P}_{n}^{\left( \alpha ,\beta \right) }\left( x\right) =\frac{\Gamma
\left( 2n+\alpha +\beta +1\right) }{2^{n}n!\Gamma \left( n+\alpha +\beta
+1\right) }x^{n}+O\left( x^{n-1}\right) ,%
\end{array}%
\right.  \label{asmptJacobi}
\end{equation}%
from Eq(\ref{foHDPT}) and Eq(\ref{fondDPT}), we deduce for $n>\beta -\alpha
-1$

\begin{equation}
\left\{ 
\begin{array}{c}
\psi _{n}\left( x;\alpha ,\beta \right) \underset{x\rightarrow 0^{+}}{\sim }%
x^{\alpha +1/2}\frac{\left( \alpha +1\right) _{n}}{n!}\underset{x\rightarrow
0^{+}}{\rightarrow }0^{+} \\ 
\psi _{n}\left( x;\alpha ,\beta \right) \underset{x\rightarrow +\infty }{%
\sim }e^{\left( 2n+\alpha -\beta +1\right) x}\frac{\left( 2n+\alpha -\beta
\right) ...\left( n+\alpha -\beta +1\right) }{2^{n}n!}\underset{x\rightarrow
+\infty }{\rightarrow }+\infty .%
\end{array}%
\right.
\end{equation}

$\psi _{n}$ having the same sign at both ends of the definition interval, $%
\psi _{n}$ has no zero on $\mathbb{R}$. The DBT built on the corresponding
RS functions generate isospectral extensions of the HDPT potential. They are
given by

\begin{equation}
V^{\left( n\right) }(x;\alpha ,\beta )=V(x;\alpha ,\beta )+w_{n}^{\prime
}\left( x;\alpha ,\beta \right) ,
\end{equation}%
where $w_{n}^{\prime }$ is given in Eq(\ref{RSHDPT}). Since $1/\psi
_{n}(x;\alpha ,\beta )$ is divergent at the origin, it is not a physical
eigenstate of $V^{\left( n\right) }$ which is then strictly isospectral to $%
V $. It admits for bound state eigenfunctions

\begin{equation}
\psi _{k}^{\left( n\right) }(x;\alpha ,\beta )=\left( w_{n}(x;\alpha ,\beta
)-w_{k}(x;\alpha ,\beta )\right) \psi _{k}(x;\alpha ,\beta )=\psi _{0}\left(
x;\alpha +1,\beta -1\right) \frac{\mathit{P}_{n,k}^{\left( \alpha ,\beta
\right) }\left( z\right) }{\mathit{P}_{n}^{\left( \alpha ,-\beta \right)
}\left( z\right) },
\end{equation}%
where $k<a$ and where

\begin{equation}
\mathit{P}_{n,k}^{\left( \alpha ,\beta \right) }\left( z\right) =\left(
k+\alpha -\beta +1\right) \mathit{P}_{n}^{\left( \alpha ,-\beta \right)
}\left( z\right) \mathit{P}_{k-1}^{\left( \alpha +1,-\beta +1\right) }\left(
z\right) -\left( n+\alpha -\beta +1\right) \mathit{P}_{n-1}^{\left( \alpha
+1,-\beta +1\right) }\left( z\right) \mathit{P}_{k}^{\left( \alpha ,-\beta
\right) }\left( z\right)
\end{equation}%
is a polynomial of degree $n+k-1$.

\subsection{Case of dispersion relation which is a second degree Laurent
polynomial}

\subsubsection{Eckart potential}

The Eckardt\bigskip\ potential ($x\in \left] 0,+\infty \right[ $) can be
written as \cite{cooper,Dutt,grandati}%
\begin{equation}
V(x;a,b)=a(a+1)y^{2}-2by+V_{0}(a,b),\ \ a^{2}<b,\ a,b>0,
\end{equation}%
where$\ y=\coth x$ and 
\begin{equation}
V(a,b)=\frac{b^{2}}{a^{2}}+a.
\end{equation}

The dispersion relation corresponds to a second degree Laurent polynomial of
the form ($a_{n}=a+n$)

\begin{eqnarray}
E_{n} &=&a^{2}+\frac{b^{2}}{a^{2}}-a_{n}^{2}-\frac{b^{2}}{a_{n}^{2}}
\label{specEc} \\
&=&-\frac{n}{\left( n+a\right) ^{2}}\left( n+2a\right) \left( n+\left( a-%
\frac{b}{a}\right) \right) \left( n+\left( a+\frac{b}{a}\right) \right) . 
\notag
\end{eqnarray}

The bound states are obtained on the increasing part of $E_{n}$, that is for 
$0\leq n<\sqrt{b}-a$ and they are given by

\begin{equation}
\psi _{n}\left( x;a,b\right) =\psi _{0}\left( x;\alpha _{n},\beta
_{n}\right) \mathit{P}_{n}^{\left( \alpha _{n},\beta _{n}\right) }\left(
y\right) ,  \label{foEc}
\end{equation}%
where

\begin{equation}
\psi _{0}\left( x;a,b\right) =\left( y-1\right) ^{\frac{\alpha }{2}}\left(
y+1\right) ^{\frac{\beta }{2}}=e^{-bx/a}\sinh ^{a}x,  \label{fondEc}
\end{equation}%
with

\begin{equation}
\left\{ 
\begin{array}{c}
\alpha _{n}=-a_{n}+\frac{b}{a_{n}} \\ 
\beta _{n}=-a_{n}-\frac{b}{a_{n}}.%
\end{array}%
\right.
\end{equation}

The corresponding RS\ functions are

\begin{equation}
w_{n}\left( x;a,b\right) =w_{0}\left( x;\alpha _{n},\beta _{n}\right)
+\left( y^{2}-1\right) \frac{n+\alpha _{n}+\beta _{n}+1}{2}\frac{\mathit{P}%
_{n-1}^{\left( \alpha _{n}+1,\beta _{n}+1\right) }\left( y\right) }{\mathit{P%
}_{n}^{\left( \alpha _{n},\beta _{n}\right) }\left( y\right) },
\label{RSfonctEc}
\end{equation}%
where

\begin{equation}
w_{0}\left( x;\alpha ,\beta \right) =\frac{\alpha }{2}\left( y+1\right) +%
\frac{\beta }{2}\left( y-1\right) .  \label{RSfondEc}
\end{equation}

Beyond the value $n=a-\sqrt{b}$, the $\psi _{n}$ do not correspond anymore
to physical eigenstates and for $n>\frac{b}{a}-a$, $E_{n}$ becoming
negative, the corresponding Schr\"{o}dinger equation enters in a
disconjugacy regime.

Using Eq(\ref{asmptJacobi}), from Eq(\ref{RSfonctEc}) we deduce the
following asymptotic behaviour for $\psi _{n}$ ($\alpha _{n}+\beta
_{n}=-2a-2n$)

\begin{equation}
\left\{ 
\begin{array}{c}
\psi _{n}\left( x;a,b\right) \underset{x\rightarrow 0^{+}}{\sim }\left(
-2a\right) ...\left( -n-2a+1\right) x^{-\left( 2n+\alpha _{n}+\beta
_{n}\right) /2} \\ 
\psi _{n}\left( x;a,b\right) \underset{x\rightarrow +\infty }{\sim }\left(
y-1\right) ^{\frac{\alpha _{n}}{2}}\frac{\left( \alpha _{n}+1\right)
...\left( \alpha _{n}+n\right) }{n!}.%
\end{array}%
\right.
\end{equation}

When $n>\frac{b}{a}-a$, $\alpha _{n}+n<0$ and $\alpha _{n}+\beta _{n}=-2a-2n$%
. Consequently

\begin{equation}
\left\{ 
\begin{array}{c}
\psi _{n}\left( x;a,b\right) \underset{x\rightarrow 0^{+}}{\rightarrow }\pm
\infty \\ 
\psi _{n}\left( x;a,b\right) \underset{x\rightarrow +\infty }{\rightarrow }%
0^{\pm },%
\end{array}%
\right.
\end{equation}%
with $\pm =\left( -1\right) ^{n}$.

Since $\psi _{n}$ has the same sign at both ends of the definition interval,
the disconjugacy of the Schr\"{o}dinger equation that it satisfies implies
that $\psi _{n}$ has no zero on $\mathbb{R}$. We can then use the
corresponding RS functions to build isospectral extensions of the Eckart
potential given by%
\begin{equation}
V^{\left( n\right) }(x;a,b)=V(y;a,b)-2\left( y^{2}-1\right) \frac{%
dw_{n}\left( y;a,b\right) }{dy},
\end{equation}%
where $w_{n}$ is given in Eq(\ref{RSfonctEc}). $1/\psi _{n}(x;a,b)$ is
divergent at $+\infty $ and is not a physical eigenstate of $V^{\left(
n\right) }$ which is then strictly isospectral to $V$. The bound state
eigenfunctions of $V^{\left( n\right) }$ are given by

\begin{equation}
\psi _{k}^{\left( n\right) }(x;a,b)=\left( w_{n}(x;a,b)-w_{k}(x;a,b)\right)
\psi _{k}(x;a,b)=\psi _{0}\left( x;\alpha _{k},\beta _{k}\right) \frac{%
\mathit{P}_{n,k}^{\left( \alpha ,\beta \right) }\left( z\right) }{2\mathit{P}%
_{n}^{\left( \alpha _{n},\beta _{n}\right) }\left( z\right) },
\end{equation}%
where $k<\sqrt{b}-a$ and

\begin{eqnarray}
\mathit{P}_{n,k}^{\left( \alpha ,\beta \right) }\left( z\right) &=&\left(
\alpha _{n}\left( y+1\right) +\beta _{n}\left( y-1\right) \right) \mathit{P}%
_{n}^{\left( \alpha _{n},\beta _{n}\right) }\left( z\right) \mathit{P}%
_{k}^{\left( \alpha _{k},\beta _{k}\right) }\left( z\right) \\
&&+\left( y^{2}-1\right) \left( n+\alpha _{n}+\beta _{n}+1\right) \mathit{P}%
_{k}^{\left( \alpha _{k},\beta _{k}\right) }\left( z\right) \mathit{P}%
_{n-1}^{\left( \alpha _{n}+1,\beta _{n}+1\right) }\left( z\right)  \notag \\
&&-\left[ k\leftrightarrow n\right] .  \notag
\end{eqnarray}

\subsubsection{Hyperbolic Rosen-Morse (HRM)\protect\bigskip\ potential}

Finally consider the HRM\bigskip\ potential ($x\in \mathbb{R}$) which is
given by \cite{cooper,Dutt,grandati}%
\begin{equation}
V(x;a,b)=a(a+1)y^{2}+2by+V_{0}(a,b),\ \ a^{2}>b,\ a,b>0,  \label{potHRM}
\end{equation}%
where$\ y=\tanh x$ and 
\begin{equation}
V(a,b)=\frac{b^{2}}{a^{2}}-a.
\end{equation}

As for the Eckart potential, the dispersion relation corresponds to a second
degree Laurent polynomial

\begin{eqnarray}
E_{n} &=&a^{2}+\frac{b^{2}}{a^{2}}-a_{n}^{2}-\frac{b^{2}}{a_{n}^{2}}
\label{specHRM} \\
&=&-\frac{n}{\left( n-a\right) ^{2}}\left( n-2a\right) \left( n-\left( a-%
\frac{b}{a}\right) \right) \left( n-\left( a+\frac{b}{a}\right) \right) 
\notag
\end{eqnarray}%
($a_{n}=a-n$), but $E_{n}$ is now singular at the positive value $n=a$.

As in the preceding cases, the bound states are obtained on the increasing
part of $E_{n}$, that is, for $0\leq n<a-\sqrt{b}$. They are given by

\begin{equation}
\psi _{n}\left( x;a,b\right) =\psi _{0}\left( x;\alpha _{n},\beta
_{n}\right) \mathit{P}_{n}^{\left( \alpha _{n},\beta _{n}\right) }\left(
y\right) ,  \label{foHRM}
\end{equation}%
where

\begin{equation}
\psi _{0}\left( x;a,b\right) =\left( 1-y\right) ^{\frac{\alpha }{2}}\left(
1+y\right) ^{\frac{\beta }{2}}=\frac{e^{-bx/a}}{\cosh ^{a}x},
\label{fondHRM}
\end{equation}%
with

\begin{equation}
\left\{ 
\begin{array}{c}
\alpha _{n}=a_{n}+\frac{b}{a_{n}} \\ 
\beta _{n}=a_{n}-\frac{b}{a_{n}}.%
\end{array}%
\right.
\end{equation}

The corresponding RS\ functions are

\begin{equation}
w_{n}\left( x;a,b\right) =w_{0}\left( x;\alpha _{n},\beta _{n}\right)
-\left( 1-y^{2}\right) \frac{n+\alpha _{n}+\beta _{n}+1}{2}\frac{\mathit{P}%
_{n-1}^{\left( \alpha _{n}+1,\beta _{n}+1\right) }\left( y\right) }{\mathit{P%
}_{n}^{\left( \alpha _{n},\beta _{n}\right) }\left( y\right) },
\label{RSfonctHRM}
\end{equation}%
where

\begin{equation}
w_{0}\left( x;\alpha ,\beta \right) =\frac{\alpha }{2}\left( 1+y\right) -%
\frac{\beta }{2}\left( 1-y\right) .  \label{RSfondHRM}
\end{equation}

Beyond the value $n=a-\sqrt{b}$, the $\psi _{n}$ do not correspond anymore
to physical eigenstates and $E_{n}$ is negative when $n$ belongs to the
intervals $\left[ a-b,a+b\right] $ and $\left[ 2a,+\infty \right] $.

Using \cite{szego,magnus}

\begin{equation}
\left\{ 
\begin{array}{c}
\mathit{P}_{n}^{\left( \alpha ,\beta \right) }\left( 1\right) =\frac{\left(
\alpha +1\right) _{n}}{n!} \\ 
\mathit{P}_{n}^{\left( \alpha ,\beta \right) }\left( -1\right) =\left(
-1\right) ^{n}\frac{\left( \beta +1\right) _{n}}{n!},%
\end{array}%
\right.  \label{limitjacobi}
\end{equation}%
from Eq(\ref{RSfonctHRM}) we deduce the following asymptotic behaviour for $%
\psi _{n}$

\begin{equation}
\left\{ 
\begin{array}{c}
\psi _{n}\left( x;a,b\right) \underset{x\rightarrow -\infty }{\sim }\left(
1+y\right) ^{\frac{\beta _{n}}{2}}\left( -1\right) ^{n}\frac{\left( \beta
_{n}+1\right) ...\left( \beta _{n}+n\right) }{n!} \\ 
\psi _{n}\left( x;a,b\right) \underset{x\rightarrow +\infty }{\sim }\left(
1-y\right) ^{\frac{\alpha _{n}}{2}}\frac{\left( \alpha _{n}+1\right)
...\left( \alpha _{n}+n\right) }{n!}.%
\end{array}%
\right.
\end{equation}

Consider first the case (i) where $a-\frac{b}{a}<n<a$. Then $a_{n}>0$, $%
\alpha _{n}>0$ and $\beta _{n}+n<0$. Consequently

\begin{equation}
\left\{ 
\begin{array}{c}
\psi _{n}\left( x;a,b\right) \underset{x\rightarrow -\infty }{\rightarrow }%
+\infty \\ 
\psi _{n}\left( x;a,b\right) \underset{x\rightarrow +\infty }{\rightarrow }%
0^{+}.%
\end{array}%
\right.
\end{equation}

Consider now the case (ii) where $a+\frac{b}{a}>n>a$ ($a_{n}<0$). In this
case we have $\alpha _{n}+n<0$ and $\beta _{n}>0$. It results

\begin{equation}
\left\{ 
\begin{array}{c}
\psi _{n}\left( x;a,b\right) \underset{x\rightarrow -\infty }{\rightarrow }%
\pm \infty \\ 
\psi _{n}\left( x;a,b\right) \underset{x\rightarrow +\infty }{\rightarrow }%
0^{\pm },%
\end{array}%
\right.
\end{equation}%
where $\pm =\left( -1\right) ^{n}$.

Finally consider the case (iii) $n>2a$ ($a_{n}<0$). We have $\alpha
_{n},\beta _{n}<0$ and $\alpha _{n}+n,\beta _{n}+n>0$. $\psi _{n}$ is then
divergent both at $+\infty $ and $-\infty $. Depending on the value of $n$
(compared to $a+b/k,$ $k\in \mathbb{N}^{\ast }$), we can then have the same
sign or not for the limits of $\psi _{n}$.

When $\psi _{n}$ has the same sign at both ends of the definition interval,
the disconjugacy of the Schr\"{o}dinger equation that it satisfies implies
that $\psi _{n}$ has no zero on $\mathbb{R}$. We can then use the
corresponding RS functions to build isospectral extensions of the HRM
potential given by%
\begin{equation}
V^{\left( n\right) }(x;a,b)=V(y;a,b)+2\left( 1-y^{2}\right) \frac{%
dw_{n}\left( y;a,b\right) }{dy},
\end{equation}%
where $w_{n}$ is given in Eq(\ref{RSfonctHRM}).

In the cases (i) and (ii), $1/\psi _{n}(x;a,b)$ is divergent at $+\infty $
or $-\infty $ and is not a physical eigenstate of $V^{\left( n\right) }$
which is then strictly isospectral to $V$.

At the contrary in the case (iii), when $1/\psi _{n}(x;a,b)$ is regular, it
also satisfies the required Dirichlet boundary conditions and constitutes
the fundamental bound state of $V^{\left( n\right) }$. The DBT is then a
backward SUSY partnership. We will not consider this case in the following.

In the cases (i) and (ii), the bound state eigenfunctions of $V^{\left(
n\right) }$ are given by

\begin{equation}
\psi _{k}^{\left( n\right) }(x;a,b)=\left( w_{n}(x;a,b)-w_{k}(x;a,b)\right)
\psi _{k}(x;a,b)=\psi _{0}\left( x;\alpha _{k},\beta _{k}\right) \frac{%
\mathit{P}_{n,k}^{\left( \alpha ,\beta \right) }\left( z\right) }{2\mathit{P}%
_{n}^{\left( \alpha _{n},\beta _{n}\right) }\left( z\right) },
\end{equation}%
where $k<a-\sqrt{b}$ and

\begin{eqnarray}
\mathit{P}_{n,k}^{\left( \alpha ,\beta \right) }\left( z\right) &=&\left(
1-y^{2}\right) \left( k+\alpha _{k}+\beta _{k}+1\right) \mathit{P}%
_{n}^{\left( \alpha _{n},\beta _{n}\right) }\left( z\right) \mathit{P}%
_{k-1}^{\left( \alpha _{k}+1,\beta _{k}+1\right) }\left( z\right) \\
&&-\left( \alpha _{k}\left( 1+y\right) -\beta _{k}\left( 1-y\right) \right) 
\mathit{P}_{n}^{\left( \alpha _{n},\beta _{n}\right) }\left( z\right) 
\mathit{P}_{k}^{\left( \alpha _{k},\beta _{k}\right) }\left( z\right)  \notag
\\
&&-\left[ k\leftrightarrow n\right] .  \notag
\end{eqnarray}%
\bigskip

\section{Shape invariance}

For the isotonic oscillator we have proven explicitely in \cite%
{grandati3,grandati5} that the shape invariance property of the initial
potential is transmitted to all its strictly isospectral ($L1$ and $L2$
series) successive extensions. This is not the case for the extensions
obtained from the potentials considered in \cite{grandati4} via the use of
regularizing symmetries.

To look for such a property for the extensions Morse and HDPT potentials
obtained above, we have to consider the superpartner of the potential $%
V^{\left( n\right) }(x;a)$ which is defined as

\begin{equation}
\widetilde{V}^{\left( n\right) }(x;a)=V^{\left( n\right) }(x;a)+2\left(
w_{0}^{\left( n\right) }(x;a)\right) ^{\prime },  \label{SUSYpart}
\end{equation}%
$w_{0}^{\left( n\right) }(x;\omega ,a)$ being the RS function associated to
the ground level of $V^{\left( n\right) }$.

Since (see Eq(\ref{transfoback2}))

\begin{equation}
w_{0}^{\left( n\right) }(x;a)=-w_{n}(x;a)-\frac{E_{n}(a)}{%
w_{n}(x;a)-w_{0}(x;a)},  \label{transfoback3}
\end{equation}%
we have, using Eq(\ref{extpot})

\begin{equation}
\widetilde{V}^{\left( n\right) }(x;a)=V(x;a)-2\left( \frac{E_{n}(a)}{%
w_{n}(x;a)-w_{0}(x;a)}\right) ^{\prime }.  \label{SUSYn}
\end{equation}

We suppose that $V$ is a TSIP satisfying

\begin{equation}
\widetilde{V}(x;a)=V(x;a)+2w_{0}^{\prime }(x;a)=V(x;a-1)+E_{1}(a).
\label{SIP}
\end{equation}

Inserting Eq(\ref{SIP}) into Eq(\ref{SUSYn}), it results

\begin{equation}
\widetilde{V}^{\left( n\right) }(x;a)=V(x;a-1)+E_{1}(a)-2W^{\prime }(x;a),
\label{SUSYpartextension}
\end{equation}%
where

\begin{equation}
W(x;a)=w_{0}(x;a)+\frac{E_{n}(a)}{w_{n}(x;a)-w_{0}(x;a)}.
\label{termeadditionnel}
\end{equation}

\subsection{\protect\bigskip Morse potential}

Consider the case of the Morse potential. Using Eq(\ref{RSfonctmorse}) and
Eq(\ref{RSfondmorse}), Eq(\ref{termeadditionnel}) gives ($z=2by$)

\begin{equation}
W(x;a)=-a_{n}+by+z\frac{\mathit{L}_{n-2}^{2a_{n}+1}(z)}{\mathit{L}%
_{n-1}^{2a_{n}}(z)}  \label{termeadditionnelmorse}
\end{equation}%
and since $a_{n}=\left( a-1\right) _{n-1}$, we deduce

\begin{equation}
W(x;a)=-\left( w_{0}(x;\left( a-1\right) _{n-1})-z\frac{\mathit{L}%
_{n}^{2\left( a-1\right) _{n-1}-1}(z)}{\mathit{L}_{n-1}^{2\left( a-1\right)
_{n-1}}(z)}\right) =-w_{n-1}(x;\left( a-1\right) _{n-1}).
\end{equation}

Consequently

\begin{eqnarray}
\widetilde{V}^{\left( n\right) }(x;a) &=&V(x;a-1)+2w_{n-1}^{\prime
}(x;\left( a-1\right) _{n-1})+E_{1}(a)  \notag \\
&=&V^{(n-1)}(x;a_{1})+E_{1}(a).
\end{eqnarray}

This is not strictly speaking a shape invariance in the sense of
Gendenshtein \cite{Gendenshtein}. As noted by Quesne \cite{quesne6}, we
rather obtain a kind of enlarged shape invariance property where the SUSY QM
partner of the n$^{th}$ extended potential $V^{(n)}$ has not the functional
form of $V^{(n)}$ (with translated parameters and an additional constant)
but the one of the preceding extension $V^{(n-1)}$.

\subsection{HDPT potential}

For the HDPT\ potential, combining Eq(\ref{RSHDPT}) and Eq(\ref{RSfondDPT})
in Eq(\ref{termeadditionnel}), we find that $W$ takes the form ($z=\cosh 2x$)

\begin{equation}
W(x;\alpha ,\beta )=w_{0}\left( x;\alpha ,\beta \right) +\frac{4n\mathit{P}%
_{n}^{\left( \alpha ,-\beta \right) }\left( z\right) }{\sinh 2x\mathit{P}%
_{n-1}^{\left( \alpha +1,1-\beta \right) }\left( z\right) }.  \label{Wjacobi}
\end{equation}

But $\mathit{P}_{n}^{\left( \alpha ,-\beta \right) }\left( z\right) $
satisfies the differential equation

\begin{equation}
\left( 1-z^{2}\right) y^{\prime \prime }(z)+\left( \alpha +\beta -z\left(
\alpha -\beta +2\right) \right) y^{\prime }(z)+n\left( n+\alpha -\beta
+1\right) y(z)=0,  \label{eqdiffjacobi}
\end{equation}%
which, combined to Eq(\ref{derivjacobi}) gives

\begin{equation}
4n\mathit{P}_{n}^{\left( \alpha ,-\beta \right) }\left( z\right) =\left(
z^{2}-1\right) \left( n+\alpha -\beta +2\right) \mathit{P}_{n-2}^{\left(
\alpha +2,-\beta +2\right) }\left( z\right) +2\left( \alpha +\beta -z\left(
\alpha -\beta +2\right) \right) \mathit{P}_{n-1}^{\left( \alpha +1,-\beta
+1\right) }\left( z\right) .\text{ }
\end{equation}

Substituting this result in Eq(\ref{Wjacobi}) and using Eq(\ref{RSfondDPT}),
we obtain

\begin{eqnarray}
W(x;\alpha ,\beta ) &=&\frac{\alpha +\beta +\left( \alpha -\beta +3\right) z%
}{\sinh 2x}+\sinh 2x\left( n+\alpha -\beta +2\right) \frac{\mathit{P}%
_{n-2}^{\left( \alpha +2,-\beta +2\right) }\left( z\right) }{\mathit{P}%
_{n-1}^{\left( \alpha +1,1-\beta \right) }\left( z\right) } \\
&=&-w_{n-1}\left( x;\alpha _{1},\beta _{1}\right) ,  \notag
\end{eqnarray}%
that is,

\begin{eqnarray}
\widetilde{V}^{\left( n\right) }(x;\alpha ,\beta ) &=&V(x;\alpha _{1},\beta
_{1})+2w_{n-1}^{\prime }(x;\alpha _{1},\beta _{1})+E_{1}(a)  \notag \\
&=&V^{(n-1)}(x;\alpha _{1},\beta _{1})+E_{1}(a).
\end{eqnarray}

Again, $V^{\left( n\right) }$ satisfies the enlarged shape invariance
property defined above.

\section{Conclusion}

For PTSIP with a finite number of bound states, we have enlarged the
generalized SUSY QM partnership presented in \cite%
{grandati,grandati2,grandati3,grandati4} by showing another way to obtain
disconjugated unphysical eigenfunctions which may serve to generate regular
rational extensions of these potentials. We have studied more explicitely
those of these extended potentials which are strictly isospectral to the
initial one. These results encompass in particular those obtained very
recently by Quesne \cite{quesne6} for the Morse potential. The enlarged
shape invariance property revealed by Quesne for the Morse potential
extensions is proven to be shared by the HDPT\ potential.

\section{Acknowledgments}

I would like to thank C.\ Quesne for very stimulating and enlightening
exchanges.


\begin{thebibliography}{0}
\expandafter\ifx\csname natexlab\endcsname\relax\def\natexlab#1{#1}\fi
\expandafter\ifx\csname bibnamefont\endcsname\relax
  \def\bibnamefont#1{#1}\fi
\expandafter\ifx\csname bibfnamefont\endcsname\relax
  \def\bibfnamefont#1{#1}\fi
\expandafter\ifx\csname citenamefont\endcsname\relax
  \def\citenamefont#1{#1}\fi
\expandafter\ifx\csname url\endcsname\relax
  \def\url#1{\texttt{#1}}\fi
\expandafter\ifx\csname urlprefix\endcsname\relax\def\urlprefix{URL }\fi
\providecommand{\bibinfo}[2]{#2}
\providecommand{\eprint}[2][]{\url{#2}}

\end{thebibliography}


\begin{thebibliography}{99}
\bibitem{cooper} F.\ Cooper, A.\ Khare and U.\ Sukhatme, \textit{%
Supersymmetry in Quantum Mechanics} (World Scientific, Singapore, 2001).

\bibitem{Dutt} R.\ Dutt, A. Khare and U.\ P.\ Sukhatme, \textquotedblleft
Supersymmetry, shape invariance and exactly solvable
potentials,\textquotedblright\ Am. J. Phys. 5\textbf{6}, 163--168 (1988).

\bibitem{Gendenshtein} L.\ Gendenshtein, \textquotedblleft Derivation of
exact spectra of the Schrodinger equation by means of
supersymmetry,\textquotedblright\ JETP Lett. \textbf{38}, 356-359 (1983).

\bibitem{gomez} D. G\'{o}mez-Ullate, N. Kamran and R. Milson,
\textquotedblleft The Darboux transformation and algebraic deformations of
shape invariant potentials\textquotedblright ,\ J. Phys. A \textbf{37},
1789-1804 (2004).

\bibitem{gomez2} D. G\'{o}mez-Ullate, N. Kamran and R. Milson,
\textquotedblleft Supersymmetry and algebraic Darboux
transformations\textquotedblright ,\ J. Phys. A \textbf{37}, 10065-10078
(2004).

\bibitem{gomez3} D. G\'{o}mez-Ullate, N. Kamran, and R. Milson,
\textquotedblleft An extended class of orthogonal polynomials defined by a
Sturm-Liouville problem\textquotedblright , J. Math. Anal. Appl. \textbf{359}%
, 352 (2009).

\bibitem{gomez4} D. G\'{o}mez-Ullate, N. Kamran and R. Milson,
\textquotedblleft An extension of Bochner's problem: exceptional invariant
subspaces\textquotedblright ,\ J. Approx. Theory \textbf{162,} 987-1006
(2010).

\bibitem{gomez5} D.G\'{o}mez-Ullate, N. Kamran and R. Milson,
\textquotedblleft Exceptional orthogonal polynomials and the Darboux
transformation\textquotedblright , J. Phys. A \textbf{43} 434016 (2010).

\bibitem{gomez6} D. G\'{o}mez-Ullate, N. Kamran and R. Milson,
\textquotedblleft On orthogonal polynomials spanning a non-standard
flag\textquotedblright , arXiv:1101.5584 (2011).

\bibitem{gomez7} D. G\'{o}mez-Ullate, N. Kamran and R. Milson,
\textquotedblleft Two-step Darboux transformations and exceptional Laguerre
polynomials \textquotedblright , J. Math. Anal. Appl. \textbf{387} 410-418
(2012).

\bibitem{quesne1} C. Quesne, \textquotedblleft Exceptional orthogonal
polynomials, exactly solvable potentials and supersymmetry\textquotedblright
,\ J. Phys. A \textbf{41}, 392001 (2008).

\bibitem{quesne} C. Quesne, \textquotedblleft Solvable rational potentials
and exceptional orthogonal polynomials in supersymmetric quantum
mechanics\textquotedblright , \textit{SIGMA} \textbf{5,} 084, 24 p (2009).

\bibitem{quesne2} B.\ Bagchi, C. Quesne and R.\ Roychoudhury,
\textquotedblleft Isospectrality of conventional and new extended
potentials, second-order supersymmetry and role of $\mathcal{PT}$
symmetry\textquotedblright , Pramana J. Phys. \textbf{73} 337-347 (2009).

\bibitem{quesne3} B.\ Bagchi and C. Quesne, \textquotedblleft An update on $%
\mathcal{PT}$-symmetric complexified Scarf II potential, spectral
singularities and some remarks on the rationally-extended supersymmetric
partners\textquotedblright , J. Phys. A \textbf{43} 305301 (2010).

\bibitem{quesne4} C. Quesne, \textquotedblleft Higher-order SUSY, exactly
solvable potentials, and exceptional orthogonal
polynomials\textquotedblright , Mod. Phys. Lett. A \textbf{26} 1843-1852
(2011).

\bibitem{quesne5} C. Quesne, \textquotedblleft Rationally-extended radial
oscillators and Laguerre exceptional orthogonal polynomials in kth-order
SUSYQM \textquotedblright , Int. J. Mod. Phys. A \textbf{26} 5337-5347
(2011).

\bibitem{odake} S. Odake and R. Sasaki, \textquotedblleft Infinitely many
shape invariant potentials and new orthogonal polynomials\textquotedblright
, Phys. Lett. B \textbf{679}, 414 - 417 (2009).

\bibitem{sasaki} S. Odake and R. Sasaki, \textquotedblleft Another set of
infinitely many exceptional (X$_{l}$) Laguerre polynomials\textquotedblright
, Phys. Lett. B \textbf{684}, 173-176 (2009).

\bibitem{ho} C-L. Ho, S. Odake and R. Sasaki, \textquotedblleft Properties
of the exceptional (X$_{l}$) Laguerre and Jacobi
polynomials,\textquotedblright\ YITP-09-70, arXiv :0912.5477[math-ph] (2009).

\bibitem{odake2} S.Odake and R. Sasaki, \textquotedblleft Infinitely many
shape invariant potentials and cubic identities of the Laguerre and Jacobi
polynomials\textquotedblright , J. Math. Phys. \textbf{51}, 053513 (2010).

\bibitem{sasaki2} R. Sasaki, S. Tsujimoto and A. Zhedanov, \textquotedblleft
Exceptional Laguerre and Jacobi polynomials and the corresponding potentials
through Darboux-Crum transformations\textquotedblright , J. Phys. A \textbf{%
43} 315204 (2010).

\bibitem{odake3} S. Odake and R. Sasaki, \textquotedblleft\ Exactly Solvable
Quantum Mechanics and Infinite Families of Multi-indexed Orthogonal
Polynomials \textquotedblright , Phys. Lett. B \textbf{702} 164-170(2011).

\bibitem{dutta} D.\ Dutta and P.\ Roy, \textquotedblleft Conditionally
exactly solvable potentials and exceptional orthogonal
polynomials\textquotedblright , J. Math. Phys. \textbf{51}, 042101 (2010).

\bibitem{grandati2} Y. Grandati and A. B\'{e}rard, \textquotedblleft
Solvable rational extension of translationally shape invariant
potentials\textquotedblright ,\ arXiv:0912.3061 (2009), to appear in the
proceedings of the Jairo Charris Seminar 2010, "\textit{Algebraic Aspects of
Darboux Transformations, Quantum Integrable Systems and Supersymmetric
Quantum Mechanics}".

\bibitem{grandati3} Y. Grandati, \textquotedblleft Solvable rational
extensions of the isotonic oscillator\textquotedblright , Ann. Phys.\textbf{%
\ 326}, 2074-2090 (2011).

\bibitem{grandati4} Y. Grandati, \textquotedblleft Solvable rational
extensions of the Morse and Kepler-Coulomb potentials\textquotedblright ,\
arXiv:1103.5023 (2011).

\bibitem{grandati5} Y.\ Grandati, \textquotedblleft Multistep DBT and
regular rational extensions of the isotonic oscillator \textquotedblright ,
arXiv:1108.4503 (2011).

\bibitem{grandati} Y. Grandati and A. B\'{e}rard, \textquotedblleft Rational
solutions for the Riccati-Schr\"{o}dinger equations associated to
translationally shape invariant potentials\textquotedblright ,\ Ann. Phys. 
\textbf{325}, 1235-1259 (2010).

\bibitem{quesne6} C. Quesne, \textquotedblleft\ Revisiting (quasi-)exactly
solvable rational extensions of the Morse potential \textquotedblright ,
arXiv:1203.1812 [math-ph] (2012).

\bibitem{szego} G. Szeg\"{o}, \textit{Orthogonal polynomials, (}American
Mathematical Society\textit{,} Providence, 1975).

\bibitem{magnus} A. Erd\'{e}lyi, W. Magnus, F. Oberhettinger and F. G.
Tricomi, \textit{Higher transcendental functions (}Mc Graw-Hill\textit{,}
New York, 1953).

\bibitem{carinena2} J.\ F.\ Cari\~{n}ena, A.\ Ramos and D.\ J.\ Fernandez,
\textquotedblleft Group theoretical approach to the intertwined
hamiltonians\textquotedblright ,\ Ann. Phys., \textbf{292}, 42-66 (2001).

\bibitem{Ramos} J.\ F.\ Cari\~{n}ena and A.\ Ramos, \textquotedblleft
Integrability of Riccati equation from a group theoretical
viewpoint,\textquotedblright\ Int. J. Mod. Phys. A, \textbf{14}, 1935-1951
(1999).

\bibitem{sukumar} C.\ V.\ Sukumar, \textquotedblleft Supersymmetric quantum
mechanics of one-dimensional systems\textquotedblright , J.\ Phys.\ A, 
\textbf{18}, 2917-2936 (1985).

\bibitem{hartman} P. Hartman, \textit{Ordinary differential equations, (}%
John Wiley\textit{,} New York, 1964).

\bibitem{coppel} W.\ A. Coppel, \textit{Disconjugacy, (}Springer\textit{,}
Berlin, 1971).

\bibitem{derr} V.\ Y.\ Derr, \textquotedblleft The theory of disconjugacy
for a second order linear differential equations\textquotedblright ,\
arXiv:0811.4636[mathCA] (2008).
\end{thebibliography}
\end{document}